# Model-Free Voltage Regulation of Unbalanced Distribution Network Based on Surrogate Model and Deep Reinforcement Learning

Di Cao, *Student Member, IEEE*, Junbo Zhao, *Senior Member, IEEE*, Weihao Hu, *Senior Member, IEEE,* Fei Ding, *Senior Member, IEEE,*Qi Huang, *Senior Member, IEEE,* Zhe Chen, *Fellow, IEEE*, Frede Blaabjerg, *Fellow, IEEE*

*Abstract*— **Accurate knowledge of the distribution system topology and parameters is required to achieve good voltage controls, but this is difficult to obtain in practice. This paper develops a model-free approach based on the surrogate model and deep reinforcement learning (DRL). We have also extended it to deal with unbalanced three-phase scenarios. The key idea is to learn a surrogate model to capture the relationship between the power injections and voltage fluctuation of each node from historical data instead of using the original inaccurate model affected by errors and uncertainties. This allows us to integrate the DRL with the learned surrogate model. In particular, DRL is applied to learn the optimal control strategy from the experiences obtained by continuous interactions with the surrogate model. The integrated framework contains training three networks, i.e., surrogate model, actor, and critic networks, which fully leverage the strong nonlinear fitting ability of deep learning and DRL for online decision making. Several single-phase approaches have also been extended to deal with three-phase unbalance scenarios and the simulation results on the IEEE 123-bus system show that our proposed method can achieve similar performance as those that use accurate physical models.**

## I. Introduction

Active distribution network (ADN) is an effective way of improving the utilization rate of distributed energy resources (DERs) since it can realize the local consumption and avoid the unnecessary power loss from long-distance transmission [1]. However, the uncertainties and intermittence of DERs bring numerous technical challenges for ADN operations. Among them, voltage regulation is one of the key issues.

To better regulate ADN voltage, various approaches have been proposed, which can be classified into two categories. From the perspective of utility, the control strategies include on-load tap changer (OLTC) [2], capacitor banks [3], and network reconfigurations. However, the upgrading of existing facilities to handle more complicated grid operations is costly. From the customer's point of view, the control strategies can be the curtailment of PV generation [4], the reactive power control of PV inverter [5], and the energy management of energy storage systems (ESSs) [6-7]. The curtailment of PV output reduces the economic benefits of customers and it cannot provide voltage support during the night. Although various approaches have been proposed for the control of ESSs, they suffer from high investment and maintenance costs, and this has not been largely deployed in today's ADN. By contrast, reactive power control of PV inverters is an economically attractive solution since it does not cause the waste of solar power with negligible extra investments. Tests in [8] demonstrate that the reactive power control of PV inverters can achieve the most optimum economy as compared to the active power curtailment of PV, distributed ESS, and OLTC controls.

To deal with the uncertainties of load demand and DERs, stochastic programming (SP) [9-10], robust optimization (RO) [11-14] are developed. SP methods require pre-sampling scenarios according to the assumed probability distribution. However, that information is difficult to obtain in practice. They also suffer from heavy computing burden. RO methods receive more and more attention in recent years due to its effectiveness in dealing with uncertainties. They achieve robust operation by constructing a solution that immunizes all possible realizations in the uncertainty set. Both SP and RO based control strategies are physical model-based. A common assumption is that the parameters and topology of ADN are accurate, which are also challenging to guarantee [15-16]. Moreover, SP and RO deal with the uncertainties of DERs and load demand by finding a predetermined solution. However, DERs can fluctuate a lot in a short time that may go beyond the pre-determined solutions. For example, the PV output may change rapidly in a few seconds due to cloud dynamics [17]. In this condition, more frequent operations of controllable devices are required to cope with the fast-changing outputs of PVs. But they have to re-compute the time-consuming optimal solutions and therefore difficult to be used for real-time decisions.

To address these issues, machine learning (ML)-based methods are developed. Among these, deep reinforcement learning (DRL) is widely used since it can directly learn from the experience obtained by interactions with the environment [18]. A multi-time scale voltage control strategy is proposed in [19] combining DRL and physics-based optimization. [20] develops a double deep Q-learning based method for the management of ESSs in a micro-grid. A constrained DRL with a continuous action search is proposed for the volt-VAR control of ADN in [21]. The DRL based approaches can learn the functional map from state to action from historical data by interactions with the environment. The learned strategies have generalizability to newly encountered situations without resolving the problem again. The decision process is very fast and suitable for real-time control. However, these methods have two disadvantages: 1) they rely on the model of ADN to calculate the reward during the training process and thus the assumption on accurate knowledge of the parameters and topology of the ADN is still there; 2) the practical ADN is three-phase unbalanced but these methods calculate the power flow using single-phase model.

To bridge these gaps, this paper proposes a model-free approach for the voltage regulation of three-phase unbalanced ADN utilizing the reactive power capability of PV inverters and static var compensator (SVC). The main contributions are:

1) The proposed approach is model-free that distinguishes with existing literature. This is achieved by strategically integrating the deep neural network (DNN) based surrogate model with the DRL algorithm. Specifically, a surrogate model is first trained in a supervised manner using recorded historical data to learn the relationship between the power injections and voltage fluctuations of each node. Then, the DRL algorithm is applied to learn an optimal control strategy from the experiences obtained by continuous interactions with the surrogate model.

2) The proposed approach can handle three-phase unbalance as compared to the widely used single-phase assumption, while controlling single-phase PV inverters and SVCs. The voltage regulation problem is first formulated as a Markov decision process (MDP) with finite time-steps and solved by a deep deterministic policy gradient (DDPG) algorithm. DDPG can learn powerful voltage regulation strategies from historical data and inform decisions based on the latest observations in real-time. The existing SP and DDQN algorithms for a single-phase system have been extended to deal with three-phase unbalance as well. The comparison results demonstrate that our model-free method achieves better control performance while being applicable for online decision makings.

The rest of this paper is organized as follows. Section II describes the mathematical model of the voltage regulation problem. In Section III, the surrogate model and control are illustrated in detail. Numerical results are discussed in Section IV. Section V concludes the paper.

## II. PROBLEM FORMULATION

### A. Problem Statement

The objective of voltage regulation is to minimize the voltage deviation of each node in the three-phase unbalanced ADN via

$$\min_{Q_{i,PV}^{\alpha}, Q_{i,SVC}^{\alpha}} F(x) = \sum_{i=1}^{n} \sum_{\alpha=a,b,c} |\sqrt{(e_i^{\alpha})^2 + (f_i^{\alpha})^2} - V_0| \quad (1)$$

s.t.

$$P_{i,PV}^{\alpha} - P_{i,Load}^{\alpha} - e_i^{\alpha} \sum_{j=1}^{n} \sum_{\beta=a,b,c} (G_{ij}^{\alpha\beta} e_j^{\beta} - B_{ij}^{\alpha\beta} f_j^{\beta})$$
$$- f_i^{\alpha} \sum_{j=1}^{n} \sum_{\beta=a,b,c} (G_{ij}^{\alpha\beta} f_j^{\beta} + B_{ij}^{\alpha\beta} e_j^{\beta}) = 0 \quad (2)$$

$$Q_{i,PV}^{\alpha} + Q_{i,SVC}^{\alpha} - Q_{i,Load}^{\alpha} - f_i^{\alpha} \sum_{j=1}^{n} \sum_{\beta=a,b,c} (G_{ij}^{\alpha\beta} e_j^{\beta} - B_{ij}^{\alpha\beta} f_j^{\beta})$$
$$+ e_i^{\alpha} \sum_{j=1}^{n} \sum_{\beta=a,b,c} (G_{ij}^{\alpha\beta} f_j^{\beta} + B_{ij}^{\alpha\beta} e_j^{\beta}) = 0 \quad (3)$$

$$V_{\min}^2 \leq (e_i^{\alpha})^2 + (f_i^{\alpha})^2 \leq V_{\max}^2 \quad (4)$$

$$\begin{cases} f_s^a - e_s^a \tan(\frac{0\pi}{180}) = 0 \\ f_s^b - e_s^b \tan(\frac{-120\pi}{180}) = 0 \\ f_s^c - e_s^c \tan(\frac{120\pi}{180}) = 0 \end{cases} \quad (5)$$

$$Q_{i,SVC,\min}^{\alpha} \leq Q_{i,SVC}^{\alpha} \leq Q_{i,SVC,\max}^{\alpha} \qquad s \in N_S \quad (6)$$

$$0 \leq P_{i,PV}^{\alpha} \leq P_{i,PV,\max}^{\alpha} \qquad g \in N_G \quad (7)$$

$$(P_{i,PV}^{\alpha})^2 + (Q_{i,PV}^{\alpha})^2 \leq (S_{i,PV}^{\alpha})^2 \qquad g \in N_G \quad (8)$$

where (1) is the objective function; $e_i^{\alpha}$ and $f_i^{\alpha}$ represent the real and imaginary components of complex voltage of phase $\alpha$ at bus $i$; $V_0$ represents the rated voltage; $Q_{i,PV}^{\alpha}, Q_{i,SVC}^{\alpha}$ are control variables, i.e., the reactive power of PV and SVC located at phase $\alpha$ of bus $i$; (2) and (3) are the active and reactive power flow constraints; $P_{i,PV}^{\alpha}$ represents PV active power connected to phase $\alpha$ of bus $i$; $P_{i,Load}^{\alpha}$ and $Q_{i,Load}^{\alpha}$ are the active and reactive power of load demand connected to phase $\alpha$ at bus $i$; $G_{ij}^{\alpha\beta}$ and $B_{ij}^{\alpha\beta}$ are the real and imaginary components of the complex admittance matrix elements; (4) denotes the constraint of voltage of each node; $V_{\min}$ and $V_{\max}$ are the lower and upper bounds of voltage at bus $i$; (5) represents the relationship between the three phases of root bus $s$; (6) denotes the constraint of the SVC; $Q_{i,SVC,\min}^{\alpha}$ and $Q_{i,SVC,\max}^{\alpha}$ are the lower and upper bounds of the reactive power of SVC connected to phase $\alpha$ at bus $i$; (7)-(8) represent the inequality constraints of the PV; Specifically, (7) is the active power constraint of PV; $P_{i,PV,\max}^{\alpha}$ denotes the rated power of PV connected to phase $\alpha$ at bus $i$. (8) shows the relationship between the active and reactive power of the PV inverter; $S_{i,PV}^{\alpha}$ represents the apparent power of PV inverter connected to phase $\alpha$ at bus $i$.

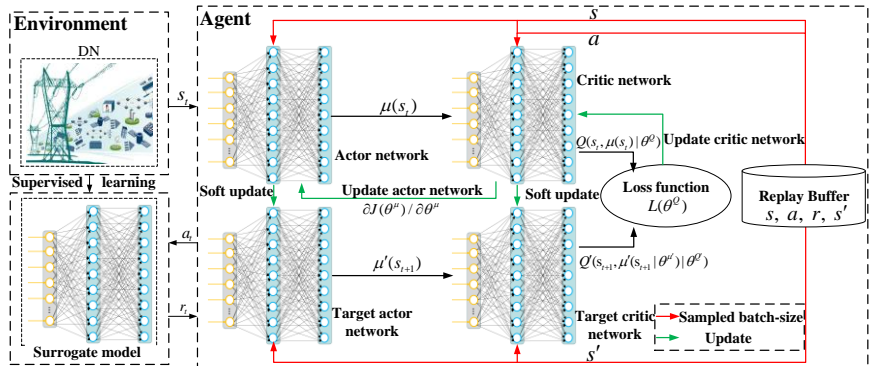

Fig. 1. Proposed model-free control framework integrating surrogate modeling and DRL.

## III. Proposed Model-Free Voltage Control Framework

The proposed model-free control framework is shown in Fig. 1. It consists of three main parts, namely the surrogate modeling, MDP formulation and DRL control.

### A. DNN for Surrogate Modeling

The aim of surrogate modeling is to learn the nonlinear maps from the active and reactive power injections to the voltage of each node. DNN is a powerful feature learning algorithm that has a strong nonlinear fitting ability. By transforming the raw input from layer to layer hierarchically, it can learn high-dimensional abstract feature representations from the training data [22]. The relationship between the voltage and active and reactive power injections can be represented as:

$$V_i = \langle W^s, g^s(P_i, Q_i) \rangle + b^s \quad (9)$$

where $\langle \cdot, \cdot \rangle$ is the inner product; $W^s$ is the weight matrix of the output layer; $b^s$ is the bias of the output layer; $V_i$, $P_i$, $Q_i$ are voltage, active and reactive power of each node at instance $i$; $g^s(\cdot)$ denotes the hierarchical transformation of the input through multi-layer nonlinear mappings. In this paper, multi fully-connected hidden layers are used. Therefore, $g^s(\cdot)$ is

$$g^s(P_t, Q_t) = g_l^s[...g_1^s(P_t, Q_t)] \quad (10)$$

$$g_i^s = \tanh(\langle W_i^s, g_{i-1}^s \rangle + b_i^s), \quad i = 2, 3, ..., l \quad (11)$$

where $g_i^s$ is the function of the $i$th layer; $W_i^s$ is the weight matrix of the $i$th layer; $b_i^s$ is the bias of the $i$th layer; $tanh$ is the activation function. The surrogate model is parameterized by $\theta^s = \{W_1^s, b_1^s, ..., W_l^s, b_l^s, W^s, b^s\}$.

### B. MDP Formulation

In this paper, the voltage regulation problem is formulated as an MDP with finite time-steps and is defined by a tuple $<S, A, R>$, where

• **State-space $S$**: the state at time slot $t$, $s_t \in S$ consists of three components: $(P_{i,Load}^\alpha, P_{i,PV}^\alpha, Q_{i,Load}^\alpha)$.

• **Action space $A$**: the action at time slot $t$, $a_t \in A$ consists of two components: $(Q_{i,PV}^\alpha, Q_{i,SVC}^\alpha)$.

• **Reward function $R$**: the objective of the model is to reduce the voltage deviations. Thus, the immediate reward of an agent at time slot $t$ is $r_t = -\sum_{i=1}^{n} \sum_{\alpha=a,b,c} |\sqrt{(e_i^\alpha)^2 + (f_i^\alpha)^2} - V_0| - \eta$, where $\eta$ is the penalty term when the voltage crosses the threshold.

One MDP is composed of a finite number of time steps. At each time slot, the agent decides the control action $a_t$ based on the observed state $s_t$, obtains an immediate reward $r_t$ and system transfer to the next state. The objective of the agent is to learn a voltage regulation strategy $a_t = f(s_t)$ to maximize the discounted cumulative reward from the current time step onward $R(s_t, a_t) = r_t + \gamma r_{t+1} + ... + \gamma^{T-t} r_T$, where $\gamma \in [0,1]$ is the discount factor to balance the future reward against the immediate reward [23].

### C. DRL Control Model

DDPG is an actor-critic framework-based algorithm, which simultaneously optimizes two functions for the problem. The policy function maps the state $(P_{i,Load}^\alpha, P_{i,PV}^\alpha, Q_{i,Load}^\alpha)$ to the desired output $(Q_{i,PV}^\alpha, Q_{i,SVC}^\alpha)$. The critic function maps state and action pairs $(P_{i,Load}^\alpha, P_{i,PV}^\alpha, Q_{i,Load}^\alpha, Q_{i,PV}^\alpha, Q_{i,SVC}^\alpha)$ to the expected cumulative reward. They are trained against each other such that the critic function better predicts the outcomes, and the policy function produces control decisions with reduced voltage deviations, see Fig. 1 for the details.

*1) Critic Function*

The critic function is also named the action-value function $Q^\pi(s_t, a_t)$, which is the expected cumulative reward when action $a_t$ is taken in state $s_t$ under policy $\pi$:

$$Q^\pi(s_t, a_t) = \mathbb{E}_{s_t \sim E, \, s_{t+1} \sim E}[r(s_t, a_t) + \gamma Q^\pi(s_{t+1}, \mu(s_{t+1}))] \quad (12)$$

where $E$ is the environment. Due to the strong non-linear fitting ability of DNN, it is used to approximate action-value function:

$$Q^\pi(s_t, a_t) = g_l^Q[...g_1^Q(s_t, a_t)] \quad (13)$$

$$g_i^Q = f(W_i^Q * o_{i-1}^Q + b_i^Q), \quad i = 2, 3, ..., l \quad (14)$$

where $(s_t, a_t)$ is the input vector and $g_i^Q$ is the function of the $i$th layer; $W_i^Q$ is the weight matrix of the $i$th layer and $b_i^Q$ is the corresponding bias; $o_{i-1}^Q$ is the output of the ($i$-1)th layer; $f$ is the activation function. The action-value function is parameterized by $\theta^Q = \{W_1^Q, b_1^Q, ..., W_l^Q, b_l^Q\}$ and can be optimized by minimizing the following loss function:

$$L(\theta^Q) = \mathbb{E}_{\mu'}[(Q(s_t, a_t | \theta^Q) - y_t)^2] \quad (15)$$

$$y_t = r(s_t, a_t) + \gamma Q(s_{t+1}, u(s_{t+1}) | \theta^Q) \quad (16)$$

The critic network is trained in a supervised learning manner and $y_t$ and $Q(s_t, a_t | \theta^Q)$ should be as close as possible.

*2) Policy Function*

The policy function maps the state to action. To deal with the dynamic environment, this paper use DNN to approximate the policy function via

$$a_t = g_l^\mu[...g_1^\mu(s_t)] \quad (17)$$

$$g_i^\mu = f(W_i^\mu * o_{i-1}^\mu + b_i^\mu), \quad i = 2, 3, ..., l \quad (18)$$

where $s_t$, $g_i^\mu$ are the input and the function of the $i$th layer, respectively. $W_i^\mu$ is the weight matrix of the $i$th layer with bias $b_i^\mu$; $o_{i-1}^\mu$ is the output of the ($i$-1)th layer; $f$ is the activation function. The policy function is parameterized by $\theta^\mu = \{W_1^\mu, b_1^\mu, ..., W_l^\mu, b_l^\mu\}$. [24] shows that the parameters of the policy function should be updated towards the gradient of $J(\theta^\mu)$, i.e.,

$$\frac{\partial J(\theta^\mu)}{\partial \theta^\mu} = \mathbb{E}_{s \sim \rho^\mu}[\nabla_{\theta^\mu} Q(s, a | \theta^Q)|_{s=s_t, u=u(s_t|\theta^\mu)}] \quad (19)$$
$$= \mathbb{E}_{s \sim \rho^\mu}[\nabla_\theta \mu_\theta(s | \theta^\mu)|_{s=s_t} \nabla_a Q(s, a | \theta^Q)|_{s=s_t, u=u(s_t|\theta^\mu)}]$$

The parameters of the policy network are adjusted in the direction that maximizes the Q value, which is the expected cumulative reward that the agent achieves at a state.

*3) Target Networks*

If the critic network is directly trained according to (15) and (16), the training process may be unstable. This is because the critic network used for approximating the action-value function $Q(s_t, a_t | \theta^Q)$ is also used for the calculation of reward. To solve this problem, DDPG introduces a target critic network and a target actor-network for the calculation of the targets $y_t$. (16) is thus rewritten as

$$y_t = r(s_t, a_t) + \gamma Q'(s_{t+1}, \mu'(s_{t+1} | \theta^{\mu'}) | \theta^{Q'}) \quad (20)$$

where $\mu'(s_{t+1} | \theta^{\mu'})$ is the target actor-network parameterized by $\theta^{\mu'}$ and $Q'(s_{t+1}, \mu'(s_{t+1} | \theta^{\mu'}) | \theta^{Q'})$ is the target critic network parameterized by $\theta^{Q'}$. The parameters of the target networks are updated by slowly tracking the online neural networks: $\theta^{\mu'} \leftarrow \tau\theta + (1-\tau)\theta^\mu$, $\theta^{Q'} \leftarrow \tau\theta + (1-\tau)\theta^Q$, where $\tau \ll 1$. The target networks constrain the changes in the target value, which can guarantee the convergence of the neural network.

*4) Replay Buffer and Exploration*

During the training process, the input data should be independent and identically distributed. For the DRL algorithm, the data are correlated with each other. To improve its stability, the experience replay mechanism is adopted. The data are continuously stored in the replay buffer, from which a batch size is uniformly sampled to train the DNN. This mechanism helps break the correlation among the sequence data [24].

The effective exploration of continuous action space is the key to train the DRL algorithm. Action space exploration strategies suitable for the voltage regulation scenario are selected via

$$\mu'(s_t) = N(\mu(s_t | \theta_t^\mu), \sigma) \quad (21)$$

where the variance $\sigma$ is set as a constant when training begins and decays at a fixed speed.

*5) Algorithm Training*

There are three sets of parameters: parameters of the surrogate model $\theta^s$, critic network $\theta^Q$ and policy network $\theta^\mu$. The training process can be divided into two steps, which are shown in Table I. In the first step, the surrogate model is trained in a supervised manner. At each epoch, batches of instances are sampled to calculate the loss according to the mean squared error:

$$L(\theta^s) = \frac{1}{B}\sum_{i=1}^{B}[(V_i - \hat{V}_i(P_i, Q_i | \theta^s))^2] \quad (22)$$

where $\hat{V}_i(\cdot)$ is the predicted value of voltage. Then, stochastic gradient descent is applied to update the parameters $\theta^s$ via

$$\theta_{t+1}^s = \theta_t^s - \lambda_s \nabla_{\theta^s} L(\theta^s) \quad (23)$$

where $\lambda_s$ is the learning rate of the surrogate model. When training is completed, parameters $\theta^s$ are fixed and the surrogate model is embedded in the environment of the DDPG algorithm for the calculation of the immediate reward $r_t$.

In the second step, the parameters of the control model are optimized. Specifically, the parameters are initialized randomly and the parameters of the target networks are copies from the online network. Then, the algorithm runs for N episodes to learn the voltage regulation strategy. One epoch corresponds to a randomly sampled day from the training set. Each epoch is divided into 24-time steps, each corresponds to an hour in the day. For each time step, the agent obtains an observation of the environment $s_t$, chooses an action $a_t$, then calculates the reward $r_t$ based on the surrogate model and the environment transfers to next state $s_{t+1}$. The transition $(s_t, a_t, r_t, s_{t+1})$ is then

TABLE I
TRAINING OF THE PROPOSED METHOD

| **Algorithm** Training of the proposed method |
|---|
| 1: Randomly initialize the parameters of surrogate model $\theta^s$ |
| 2: For epoch =1, 2,…, M do |
| 3:    Sample batch from the training set $\{P_k, Q_k, V_k\}_{k=1}^B$ |
| 4:    Optimize $\theta^s$ according to equations (22) and (23) |
| 5: End for |
| 6: Fix surrogate model parameters $\theta^s$ |
| 7: Randomly initialize critic network $Q(s, a| \theta^Q)$ and actor-network $u(s | \theta^u)$ with weights $\theta^Q$ and $\theta^u$ |
| 8: Initialize target network $Q'$ and $u'$ with weights $\theta^{Q'} \leftarrow \theta^Q$, $\theta^{\mu'} \leftarrow \theta^\mu$ |
| 9: For episode =1,2,…, N do |
|       Receive initial observation $s_1$ |
|       For $t$=1,2,…24 do |
| 10:    Choose action $a_t$ according to $a_t = g_{\theta_n^\mu}^l[...g_{\theta_1^\mu}^1(s_t)] + N(\mu(s_t | \theta_t^\mu), \sigma)$, execute the action $a_t$ and transfer to the next state $s_{t+1}$, and finally calculate reward $r_t$ based on surrogate model |
| 11:    Store transition $(s_t, a_t, r_t, s_{t+1})$ in the replay buffer |
| 12:    If the replay buffer is full: $\sigma \leftarrow \sigma * \xi$ |
| 13:    Sample a random mini-batch of transitions from The replay buffer |
| 14:    Update the critic-network via equations (25)-(26) |
| 15:    Update the actor-network via equations (27)-(28) |
| 16:    Update the target actor and critic networks through $\theta^{Q'} \leftarrow \tau\theta^Q + (1-\tau)\theta^{Q'}$, $\theta^{\mu'} \leftarrow \tau\theta^\mu + (1-\tau)\theta^{\mu'}$ |
| 17: End for |
| 18: End for |

stored in the memory buffer. The actions are chosen according to an exploration strategy with Gaussian noise:

$$a_t = \mu(s_t | \theta_t^\mu) + N(\mu(s_t | \theta_t^\mu), \sigma) \quad (24)$$

In the beginning, $\sigma$ is a constant. When the memory capacity reaches the upper limit, $\sigma$ attenuates at a fixed speed. At the same time, *n* batches of experiences $(s_j, a_j, r_j, s_{j+1})$, $j$=1, 2, …, *n* are randomly sampled from the memory to update $\theta^Q$ and $\theta^\mu$. In particular, $\theta^Q$ is updated by minimizing the following loss:

$$L(\theta^Q) = \frac{1}{N}\sum_{i=1}^{N}[(Q(s_t, a_t | \theta^Q) - y_t)^2] \quad (25)$$

$\theta^Q$ is optimized by gradient descent:

$$\theta_{t+1}^Q = \theta_t^Q - \lambda_Q \nabla_{\theta^Q} L(\theta^Q) \quad (26)$$

where $\lambda_Q$ is the learning rate of the critic network. $\theta^\mu$ is updated according to the policy gradient:

$$\nabla_\mu J(\mu) = \frac{1}{N}\sum_{i=1}^{N}[\nabla_{\theta^\mu} \mu(s | \theta^\mu)|_{s=s_i} \nabla_a Q(s, a | \theta^Q)|_{s=s_i, a=u(s_i|\theta^\mu)}] \quad (27)$$

$$\theta_{t+1}^{\mu} = \theta_t^{\mu} - \lambda_{\mu} \nabla_{\theta^{\mu}} L(\theta^{\mu}) \qquad (28)$$

where $\lambda_{\mu}$ is the learning rate of the policy network. After that, the parameters in target networks are updated by slowly tracking $\theta^Q$ and $\theta^{\mu}$. When the training process is completed, the parameters of the actor network are used for voltage regulation.

## IV. NUMERICAL RESULTS

Simulations are carried out on the unbalanced IEEE 123-bus system to evaluate the performance of the proposed method. Comparative results with various benchmark methods are also provided to illustrate the advantages of the proposed method.

### A. Experimental Setup

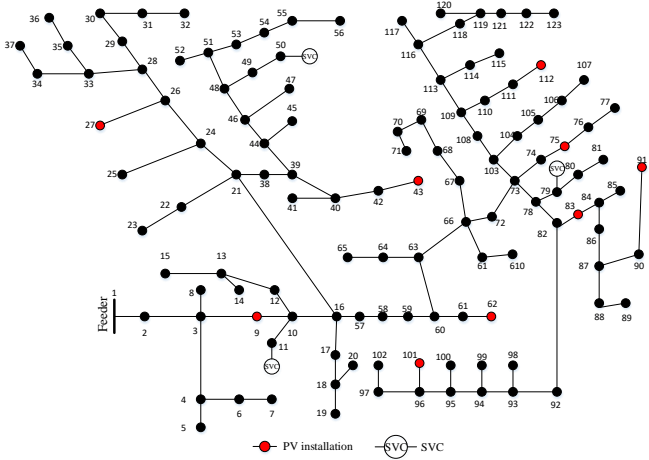

Fig. 2. Schematic of the IEEE 123-bus test system.

The schematic of the IEEE 123-bus system is shown in Fig. 2. To simulate high PV penetration, 9 PVs are connected to the ADN. The specifications of the installed PVs and SVCs are listed in Table II. The maximum voltage deviation is set as ±5% of its nominal value, yielding the upper and lower bounds as 1.05 p.u. and 0.95 p.u., respectively. One-year data of PV generations of Xiaojin, a county of China is used in this paper. The data are divided into two sets: the training set is used to train the agent to learn a voltage regulation control strategy, while the test set is applied to evaluate the performance of the learned control strategy.

TABLE II
SPECIFICATIONS OF THE CONTROLLABLE DEVICES

| Type | Capacity | Locations |
|---|---|---|
| PV | 0.6MWh/0.66MVA | 9, 27, 43, 62, 75, 83, 91, 101, 112 |
| SVC | 0.3MVar | 11, 50, 79 |

The proposed method has a surrogate model and a control model. The surrogate model is trained in a supervised manner to learn the mapping between the voltage and the active and reactive power of the nodes. 12000 instances of data composed of $\{P_i, Q_i, V_i\}_{i=1,2,...,N}$ are generated by the three-phase AC power flow model. The data are divided into two parts: 10000 instances of data are used to train the surrogate model, while the rest is used as the test set to investigate the performance of the model. A DNN with two hidden layers is used. The parameter settings of the surrogate model are shown in Table III. The control model is composed of an actor-network and a critic network, both of which share the same architecture. The parameter settings of the control model are shown in Table IV. The proposed method is written in Python with Keras. A workstation with an Intel Xeon E5-2630 CPU and an NVIDIA GeForce 1080Ti GPU is used for the training procedure.

TABLE III
PARAMETERS SETTING OF THE SURROGATE MODEL

| Parameter | Value |
|---|---|
| Neuron numbers of hidden layers | 400/400 |
| Batch size for updating NN | 32 |
| Learning rate | 0.0001 |

TABLE IV
PARAMETERS SETTING OF THE DRL METHOD

| Parameter | Value |
|---|---|
| Neuron numbers of hidden layers | 400/200 |
| Batch size for updating NN | 256 |
| Step size of each episode | 24 |
| Discount factor | 0 |
| Learning rate for actor-network | 0.001 |
| Learning rate for critic-network | 0.002 |

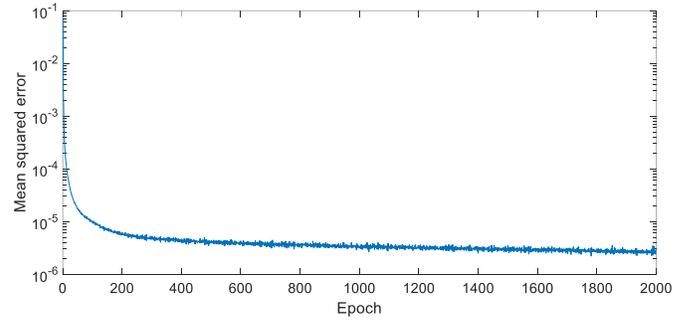

Fig. 3. The MSE of each epoch during the training process.

### B. Performance Evaluation of the Surrogate Model

For surrogate model training, the mean square error (MSE) is used as the loss function to optimize the parameters of the DNN. The loss variations with epoch during the training process are shown in Fig. 3. The MSE is shown in log scale for better visualization. The surrogate model is trained for 2000 epochs and it can be observed that the loss decreases significantly during the training process, which demonstrates that the proposed approach can learn the complex mapping relationships. Simulations are carried out on test data to evaluate the performance of the learned model. The mean absolute error (MAE) is used as the evaluation index, which is defined as follows:

$$MAE = \frac{1}{M \cdot N} \sum_{m=1}^{M} \sum_{i=1}^{N} |\hat{v}_{m,i} - v_{m,i}| \qquad (31)$$

where $M$ is the number of total instances in the test set; $N$ is the number of nodes in ADN; $\hat{v}_{m,i}$ and $v_{m,i}$ are the predicted voltage by the learned surrogate model and the true voltage of node $i$ at the $m$th instance in the test set, respectively. The MAE index represents the average prediction error of voltage at one node. The MAE on the test set is 8e-4, whose error is negligible and this shows that the prediction voltage is very close to the true voltage. The distribution of the prediction error of each node on the test set is shown in Fig. 4. It can be observed that the maximum prediction error by the developed surrogate model is 5e-3.

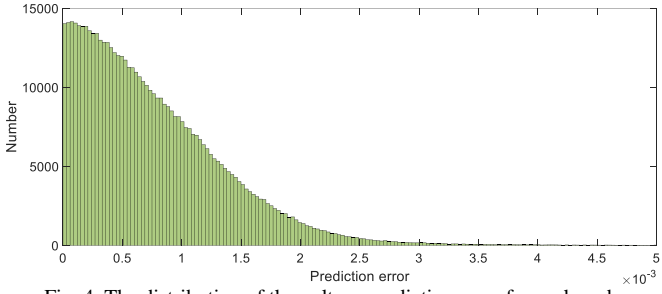

Fig. 4. The distribution of the voltage prediction error for each node.

## C. Performance Evaluation of Voltage Control

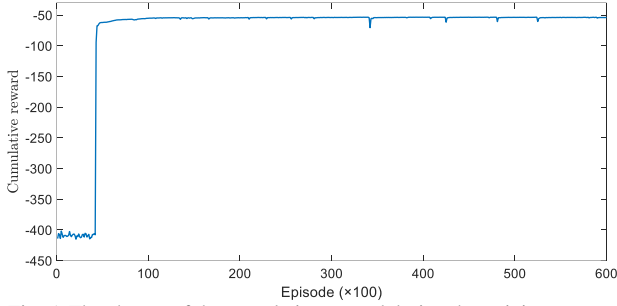

Fig. 5. The change of the cumulative reward during the training process.

TABLE V
VOLTAGE DEVIATION OF VARIOUS METHODS

| Voltage dev. | | Original | DDQN | SP | Pro. | DDPG |
|---|---|---|---|---|---|---|
| Avg. | | 3.63% | 1.51% | 0.88% | 0.82% | 0.81% |
| Avg. Dev. | a | 3.32% | 1.31% | 0.98% | 0.91% | 0.88% |
| | b | 4.16% | 1.84% | 0.72% | 0.65% | 0.63% |
| | c | 3.48% | 1.43% | 0.92% | 0.89% | 0.90% |
| Max. drop | a | 3.2% | 4.67% | 4.34% | 4.26% | 4.26% |
| | b | 2.23% | 2.46% | 1.95% | 1.67% | 1.89% |
| | c | 1.96% | 5.14% | 3.58% | 3.48% | 3.45% |
| Max. rise | a | 13.01% | 4.57% | 4.57% | 4.57% | 4.57% |
| | b | 10.69% | 5.12% | 4.57% | 4.57% | 4.57% |
| | c | 13.9% | 4.58% | 4.57% | 4.57% | 4.57% |
| Parameters dependency | | - | ✓ | ✓ | × | ✓ |

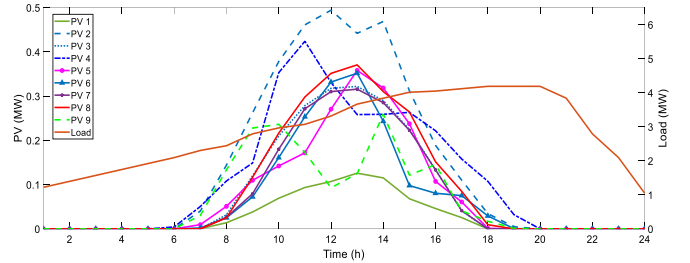

Fig. 6. The PV generations and load demand for a sunny day.

The training set is used to train the proposed control model. The value of cumulative reward during the training process is shown in Fig. 5 with 60000 episodes. It can be observed that the cumulative reward increases significantly during the training process, which demonstrates that it can learn a good policy to reduce voltage deviation by interacting with surrogate model.

To test the performance of the learned control strategy from training data, comparative tests are carried out on the test set, which consists of 30 days' data. The voltage profiles using different approaches are shown in Table V.

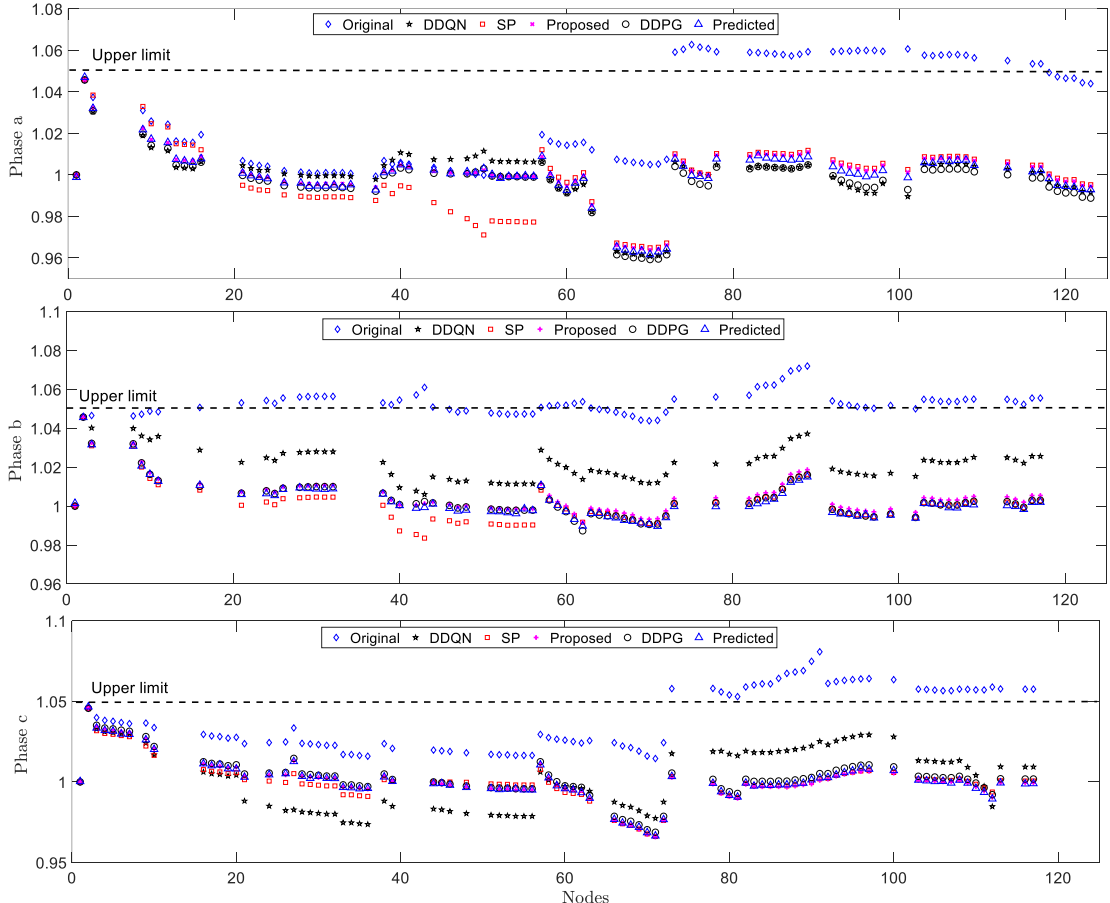

Fig. 7. Voltage profiles of all nodes before and after optimization when $t$=13:00.

The methods for comparisons include: 1) **original method** without control; 2) **double deep Q-learning (DDQN) based approach** [26]. Note that for the Q-learning based algorithm, the actions of various controllable devices must be aggregated to avoid the curse of dimensionality. The action set of the DDQN algorithm contains four variables, which control the reactive power of PVs 1-3, PVs 4-6, PVs 7-9, and the SVCs, respectively. Each variable is discretized into four values, yielding 256 actions in total. There are two shallow layers of the DNN, the number of neurons for which are 400 and 400, respectively; 3) **SP method** [9], where the PV outputs and load demand are assumed to be subject to a normal distribution, the mean standard deviations of which are 5% and 4%, respectively. 200 sets of scenarios are generated by Monte Carlo sampling, which is then reduced to 20 representative scenarios; 4) **DDPG method** [24], where the Z-bus method [25] is used to calculate the immediate reward instead of the trained surrogate model during the training of the DRL. *It is worth noting that these 4 methods have been applied for a single-phase system but in this paper, they have been extended to the three-phase system so that a fair comparison can be performed.*

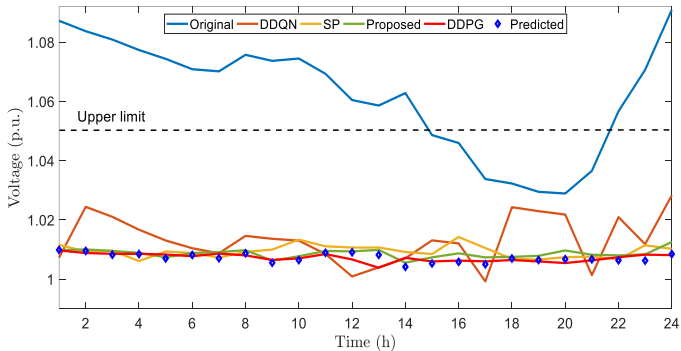

Fig. 8. The voltage profile of node 96 of phase a before and after optimization.

From Table V, it can be found that when no reactive power control is applied, the maximum voltage rise and drop are beyond the bounds. When DDQN and SP methods are used, the voltage deviation problem can be suppressed. However, the DDQN method cannot fully utilize the reactive power capability of controllable devices because of the aggregation and discretization of actions. Compared with the SP method, the proposed method and the DDPG can achieve better performance since the control decisions are made based on the latest observation instead of the generated scenarios. In addition, they do not need any assumptions on the distributions of load demand and PV generations. It is worth noting that the calculation of the control decisions for SP and the training procedures of the DDQN and DDPG depend on the exact knowledge of the parameters and topology of ADN, which are difficult to obtain in practice. By contrast, the proposed approach can obtain the control performance that is close to that of the DDPG method without the dependency on the parameters of the ADN by integrating the developed surrogate model.

A sunny day is selected as a case study to further evaluate the performance of the proposed approach. The PV generations and the load demands of the selected day are shown in Fig. 6. The voltage profiles of all nodes achieved by various methods at *t*=13:00 are shown in Fig. 7. It can be observed that the proposed approach and the DDPG method can achieve better control performance than the DDQN and SP based methods, see buses 21-56 of phases *a* and *b* for example. When the control decisions by the proposed method are implemented, the predicted voltages by the surrogate model are very close to the real value, demonstrating its effectiveness. Thanks to the good forecasting accuracy of the surrogate model, the control strategy learned by interacting with it is very similar to that of the DDPG, which requires the accurate line parameters and topology of ADN to calculate the reward value during the training process. The voltage profile of node 96 of phase *a*, which suffers from serious over-voltage problem, is shown in Fig. 8. The results are consistent with those observed in Table V and Fig. 7, demonstrating the effectiveness of the proposed approach.

### D. Robustness to Large Stochasticity

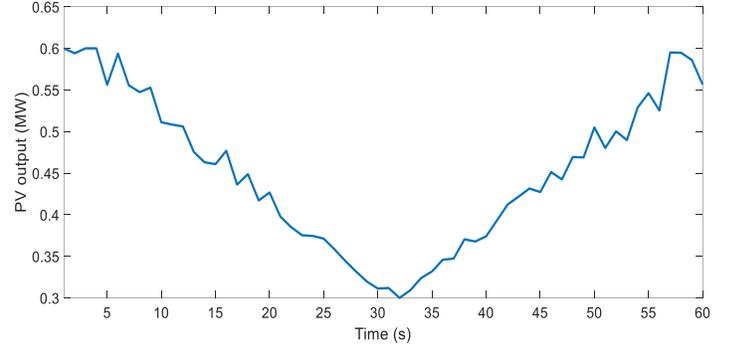

Fig. 9. The PV output profile in the dynamic simulation study.

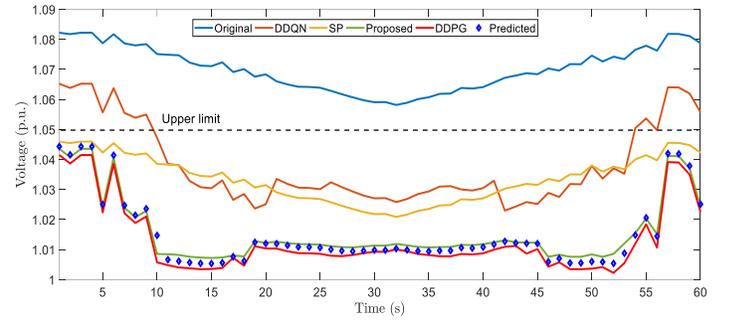

Fig. 10. The voltage profile of node 102 of phase b before and after optimization in the dynamic simulation study.

More simulations are carried out to demonstrate the advantages of the proposed method in dealing with large PV stochasticity. A rapid varying PV output in 1 min due to the cloud dynamic is tested and the PV output profile is shown in Fig. 9. In this study, the PV output starts to drop from 0.6 MW to 0.3 MW in the 30s due to the cloud dynamic. Then it starts to rise and takes 30s to go back to 0.6 MW. The voltage profile of node 102 of phase *b* before and after optimization is shown in Fig. 10. For the SP method, a pre-determined control decision is used for the voltage control of the whole process. For the DDQN method, the proposed method, and the DDPG method, the control decisions are made every second. It can be observed from the figure that when no control is used for the scheduling of reactive power, there is an over-voltage issue. When the DDQN is applied, the over-voltage problem still exists. This is because the discretization and aggregation of actions hinder the utilization of reactive power capability of controllable devices. When the SP method is applied, the over-voltage is suppressed. However, due to the fact that the predetermined control decisions cannot provide flexible reaction to the fast-changing PV output, it suffers from high voltage deviation as compared to the proposed method and the DDPG. Since the proposed approach and the DDPG method can make decisions in milliseconds, they can provide more flexible control decisions based on the latest observation and achieve better voltage regulation performance under large PV output fluctuations. It should be highlighted here that DDPG is based on the perfect ADN model while our proposed method relies on the surrogate model and DRL algorithm for control. According to the results, we can conclude that although the surrogate model approximations are applied, our proposed method still achieves quite similar performance as that of DDPG. This means that even without the accurate physical ADN network parameters and topology, our method can be applied in practice. This is the key contribution of this paper and it distinguishes from the existing methods.

## V. CONCLUSIONS

This paper proposes a model-free approach for voltage regulation of three-phase unbalanced ADN when system parameters and topology are unknown. The proposed approach consists of two main layers, namely a surrogate model and a DRL control module. The surrogate model is first trained in a supervised manner to learn the complex relationship between the voltage and active and reactive power injections of each node. Then the DRL algorithm is used to learn the voltage regulation strategy from historical data, guided by the immediate reward provided by the surrogate model. The proposed integrated method can provide voltage control in real-time without the knowledge of system parameters and topology. Comparative tests demonstrate that the proposed approach can provide a more flexible voltage regulation strategy than the predetermined control strategies by SP; it achieves better voltage regulation performance than the DDQN; it has similar performance as the DDPG that relies on accurate system information.